\documentclass{PoS}
\usepackage{tikz-feynman}
\usepackage{amsmath,braket}

\title{Form factors for the decay processes $B_c^+ \to D^0 \ell^+ \nu_{\ell}$ and $B_c^+ \to D_s^+ \ell^+ \ell^-$ from lattice QCD}

\ShortTitle{Form factors for $B_c^+ \to D^0 \ell^+ \nu_{\ell}$ and $B_c^+ \to D_s^+ \ell^+ \ell^-$}
  
\author{\speaker{Laurence Cooper}, Christine Davies\\
SUPA, School of Physics and Astronomy, University of Glasgow, Glasgow G12 8QQ, UK\\
E-mail: \email{Laurence.Cooper@glasgow.ac.uk, Christine.Davies@glasgow.ac.uk}}

\author{Matthew Wingate\\
	Department of Applied Mathematics and Theoretical Physics, University of Cambridge, Cambridge, CB3 0WA, UK\\
	E-mail: \email{M.Wingate@damtp.cam.ac.uk}}
                
\author{HPQCD Collaboration}

\abstract{We present results of the first lattice QCD calculations of the  weak matrix elements for the decays $B_c^+ \to D^0 \ell^+ \nu_{\ell}$, $B_c^+ \to D_s^+ \ell^+ \ell^-$ and $B_c^+ \to D_s^+ \nu \overline{\nu}$. Form factors across the entire physical $q^2$ range are then extracted and extrapolated to the continuum limit with physical quark masses.
Results are derived from correlation functions computed on MILC collaboration gauge configurations with three different lattice spacings and including 2+1+1 flavours of sea quarks in the Highly Improved Staggered Quark (HISQ) formalism. HISQ is also used for all of the valence quarks.
The uncertainty on the decay widths from our form factors for $B_c^+ \to D^0 \ell^+ \nu_{\ell}$ is similar in size to that from the present value for $V_{ub}$.
We obtain the ratio $\Gamma (B_{c}^{+} \rightarrow D^0 \mu^{+} \nu_{\mu}) /\left|\eta_{\mathrm{EW}} V_{u b}\right|^{2}=4.43(63) \times 10^{12} \mathrm{~s}^{-1}$. 
Combining our form factors with those found previously by HPQCD for $B_{c}^{+} \rightarrow J / \psi \mu^{+} \nu_{\mu}$, we find $\left|V_{cb}/V_{ub} \right|^2 \Gamma( B_c^+ \to D^0 \mu^+ \nu_\mu )/\Gamma(B_{c}^{+} \rightarrow J / \psi \mu^{+} \nu_{\mu}) = 0.257(36)_{B_c \to D}(18)_{B_c \to J/\psi}$.
We calculate the differential decay widths of $B_c^+ \to D_s^+ \ell^+ \ell^-$ across the full $q^2$ range, and give integrated results in $q^2$ bins that avoid possible effects from charmonium and $u \overline{u}$ resonances.
For example, we find that the ratio of differential branching fractions integrated over the range $q^2 = 1 \; \mathrm{GeV}^2 - 6 \; \mathrm{GeV}^2$ for $B_c^+ \to D_s^+ \mu^+ \mu^-$ and $B_{c}^{+} \rightarrow J / \psi \mu^{+} \nu_{\mu}$ is $6.31{\tiny }(90)_{B_c \to D_s}(65)_{B_c \to J/\psi} \times 10^{-6}$.
We also give results for the branching fraction of $B_c^+ \to D_s^+ \nu \overline{\nu}$.}

\FullConference{
	The 38th International Symposium on Lattice Field Theory, LATTICE2021 26th-30th July, 2021 \\ Zoom/Gather@Massachusetts Institute of Technology}

\graphicspath{{./}}
\begin{document}

\section{Introduction}

We use lattice QCD methods to calculate the form factors that capture the non-perturbative physics of the pseudoscalar $B_c^+$ meson decaying weakly into either $D^0 \ell^+ \nu_{\ell}$ or $D_s^+ \ell^+ \ell^-$.
This is the first time that these calculations have been performed.
More details can be found in~\cite{Cooper:2021ofu}.

We calculate the form factors $f_{0}$ and $f_+$ for $B_c^+ \to D^0 \ell^+ \nu_{\ell}$ throughout the entire range of physical momentum transfer squared, $q^2$.
An accurate prediction from the Standard Model of the normalisation and shape of these form factors will complement observations of this process from experimeant and ultimately lead to a new exclusive determination of the CKM matrix element $|V_{ub}|$ in the future.
LHCb expect~\cite{LHCbImplications} that Upgrade II will make possible a measurement of $B_{c}^{+} \rightarrow D^{0} \mu^{+} \nu_{\mu}$ with sufficient accuracy to offer a competitive determination of $V_{ub}$.
We also calculate the ratio of branching fractions for $B_c^+ \to D^0 \ell^+ \nu_{\ell}$ and $B_c^+ \to J/\psi \ell^+ \nu_{\ell}$ using the form factors from~\cite{Harrison:2020gvo}.
This allows the combination $V_{ub}/ V_{cb}$ to be examined given experimental information on this ratio.

Alongside our calculation of the form factors for $B_c^+ \to D^0 \ell^+ \nu_{\ell}$, we also carry out a lattice QCD calculation of the physical-continuum form factors $f_{0,+,T}$ for the vector and tensor current matrix elements of the rare process $B_c^+ \to D_s^+ \ell^+ \ell^-$.
The semileptonic decay $B_c^+ \to D_s^+ \ell^+ \ell^-$ is an example of a flavour-changing, neutral current (FCNC) process 
which is not allowed at tree-level in the Standard Model, thus contributions from physics beyond the Standard Model may be more visible than with tree-level decays.



The form factors calculated here are part of an ongoing programme by HPQCD to study weak decays of mesons containing a bottom quark.
We use the Highly Improved Staggered Quark formalism (HISQ)~\cite{Follana:2006rc} that is specifically designed to have small discretisation errors.
We simulate with bottom quarks at their physical mass on our finest lattice, and unphysically light bottom quarks on the coarser lattices.
Together this data informs the limit of vanishing lattice spacing and physical quark masses through HPQCD's \lq heavy-HISQ\rq~strategy. 
Recent calculations that have established the method for determining semileptonic form factors include~\cite{McLean:2019sds,McLean:2019qcx,Cooper:2020wnj,Harrison:2020gvo,Parrott:2020vbe,Harrison:2021tol}.


We also investigate strategies for improving on this first calculation of the form factors for $B_c \to D$ and $B_c \to D_s$.
These methods will inform the strategy for other future calculations of heavy-to-light quarks decays.
To minimise cost, we trial these improvements in the $B_c \to D_s$ case only and details can be found in~\cite{Cooper:2021ofu}.

\section{Lattice Methodology}

\subsection{Lattice Parameters}
Ensembles with $2+1+1$ flavours of HISQ sea quark generated by the MILC collaboration \cite{Bazavov:2010ru,Bazavov:2012xda,Bazavov:2015yea} are described in Table \ref{LattDesc1}.
The Symanzik improved gluon action used is that in \cite{Hart:2008sq} where the gluon action is improved perturbatively through $\mathcal{O}(\alpha_s a^2)$ including the effect of dynamical HISQ sea quarks.
HISQ \cite{Follana:2006rc} is used for all other valence flavours.
Our calculations feature physically massive strange quarks and equal-mass up and down quarks, with a mass denoted by $m_l$, with $m_l/m_s = 0.2$ and also the physical value $m_l/m_s = 1/27.4$ \cite{Bazavov:2014wgs}.

\begin{table*}
	\caption{Parameters for the MILC ensembles of gluon field configurations. The lattice spacing $a$ is determined for the Wilson flow parameter $w_0$~\cite{Borsanyi:2012zs}. 
		The physical value $w_0 = 0.1715(9) \; \mathrm{fm}$ was fixed from $f_{\pi}$ in \cite{Dowdall:2013rya}. 
		Sets 1 and 2 have $a \approx 0.09 \; \mathrm{fm}$.
		Set 3 has $a \approx 0.059 \; \mathrm{fm}$ and set 4 has  $a \approx 0.044\; \mathrm{fm}$.
		Sets 1, 3 and 4 have unphysically massive light quarks such that $m_l/m_s = 0.2$.
		In the fifth column, we give $n_{\mathrm{cfg}}$, the number of configurations used for each set.
		We also use four different positions for the source on each configuration.
	}
	\begin{center}
		\begin{tabular}{c c c c c c c c c c c c c} 
			\hline\hline
			set & handle & $w_0/a$ & $N_x^3 \times N_t$ & $n_\mathrm{cfg}$ & $am_l^{\mathrm{sea}}$ & $am_s^{\mathrm{sea}}$ & $am_c^{\mathrm{sea}}$ \\ [0.1ex] 
			\hline
			1 & fine & 1.9006(20) & $32^3 \times 96$ & $500$ & $0.0074$ & $0.037$ & $0.440$\\
			2 & fine-physical & 1.9518(17) & $64^3 \times 96$ & $500$ & $0.00120$ & $0.0364$ & $0.432$ \\
			3 & superfine & $2.896(6)$ & $48^3 \times 144$ & $250$ & $0.0048$ & $0.024$ & $0.286$ \\
			4 & ultrafine & $3.892(12)$ & $64^3 \times 192$ & $250$ & $0.00316$ & $0.0158$ & $0.188$ \\
			\hline\hline
		\end{tabular}
	\end{center}
	\label{LattDesc1}
\end{table*}

We work in the frame where the $B_c^+$ is at rest, and momentum is inserted into the strange and up valence quarks through partially twisted boundary conditions \cite{Sachrajda:2004mi} in the $(1\hspace{1mm}1\hspace{1mm}1)$ direction.
Values for the twists we use can be found in~\cite{Cooper:2021ofu}.
We use heavy quark masses up $am_h = 0.5, 0.65$ and $0.8$ on each set.


\subsection{Extracting the form factors}
Our calculation uses HISQ quarks exclusively.
In particular, since we use HISQ for both the parent heavy quark and the daughter light or strange quark, we can use the Partially Conserved Vector Current Ward identity to relate matrix elements of the renormalised local vector current $Z_V V^{\mu}_{\mathrm{local}}$ with matrix elements of the local scalar density through
\begin{eqnarray}
	q_{\mu} \langle D_{l(s)} | V^{\mu}_{\mathrm{local}} | H_c \rangle Z_V = (m_h - m_{l(s)})  \langle D_{l(s)} | S_{\mathrm{local}} | H_c \rangle. \label{PCVClatt}
\end{eqnarray}
This holds since the mass and scalar density multiplicative renormalisation factors $Z_m$ and $Z_S$ satisfy $Z_{m} Z_{S}=1$.
Using Eq.~\eqref{PCVClatt} to determine $Z_V$ is a fully non-perturbative strategy.
Up to discretisation effects, the renormalisation factor is independent of $q^2$, and so it is sufficient to deduce its value at zero-recoil ($\vec{q} = \boldsymbol{0}$ and maximum $q^2$).

The tensor form factor is obtained through
\begin{eqnarray}
	f_T^s \big(q^2 \big) = \frac{ Z_T \braket{D_s | T^{1,0}_{\mathrm{local}} | H_c} (M_{H_c} + M_{D_s}) }{2i M_{H_c} p_2^1}, \label{fTextract}
\end{eqnarray}
where $T^{1,0}_{\mathrm{local}}$ is the local tensor operator and $Z_T$ is its multiplicative renormalisation factor that takes the lattice tensor current to the $\overline{\mathrm{MS}}$ scheme.
We use values of the associated multiplicative renormalisation factor $Z_T$ obtained using the RI-SMOM intermediate scheme~\cite{Hatton:2020vzp}.
Values in the RI-SMOM scheme at scale $3 \; \mathrm{GeV}$ are converted to scale $m_b$ (taken as $4.8 \; \mathrm{GeV}$) in the $\overline{\mathrm{MS}}$ scheme.
Nonperturbative (condensate) artefacts in $Z_T$ in the RI-SMOM scheme were removed using analysis of the $J/\psi$ tensor decay constant~\cite{Hatton:2020vzp} .

\subsection{Fitting the form factors} \label{sec:fit_ffs}

From fits to our 3-point correlation functions, we obtain matrix elements from which we determine the form factors required.
The form factor data at all momenta and heavy quark masses on all sets in Table~\ref{LattDesc1} are then fit simultaneously to a functional form that allows for discretisation effects, dependence on the heavy meson mass, and any residual mistuning of the light, strange and charm quark bare mass parameters.

It is convenient, and now standard, to map the semileptonic region $m_{\ell}^2 <q^2< t_- = (M_{H_c} - M_{D_{l(s)}})^2$ to a region on the real axis within the unit circle through
\begin{align} \label{eqn:BcD_littlez}
	z(q^2) &= \frac{\sqrt{t_+ - q^2} - \sqrt{t_+ - t_0}}{\sqrt{t_+ - q^2} + \sqrt{t_+ - t_0}}.
\end{align}
{\tiny {\tiny }}The parameter $t_+$ is chosen to be the threshold in $q^2$ for meson pair production with quantum numbers of the current~\cite{Boyd:1997qw}, i.e. $(M_{H} + M_{\pi (K)})^2$.
We choose the parameter $t_0$ to be $0$ so that the points $q^2 = 0$ and $z = 0$ coincide.
We fit our form factors to a fit form of a truncated power series in $z$ multiplied by the factor $(1-q^2/M^2_{\text{res}})^{-1}$ which describes the dominant pole structure.
More details of the fit form we use can be found in~\cite{Cooper:2021ofu}.

\section{Results} \label{sec:results}

\subsection{Form factors} \label{sec:results_ffs}

%

\begin{figure}
	\centering
	\includegraphics[width=1.0\textwidth]{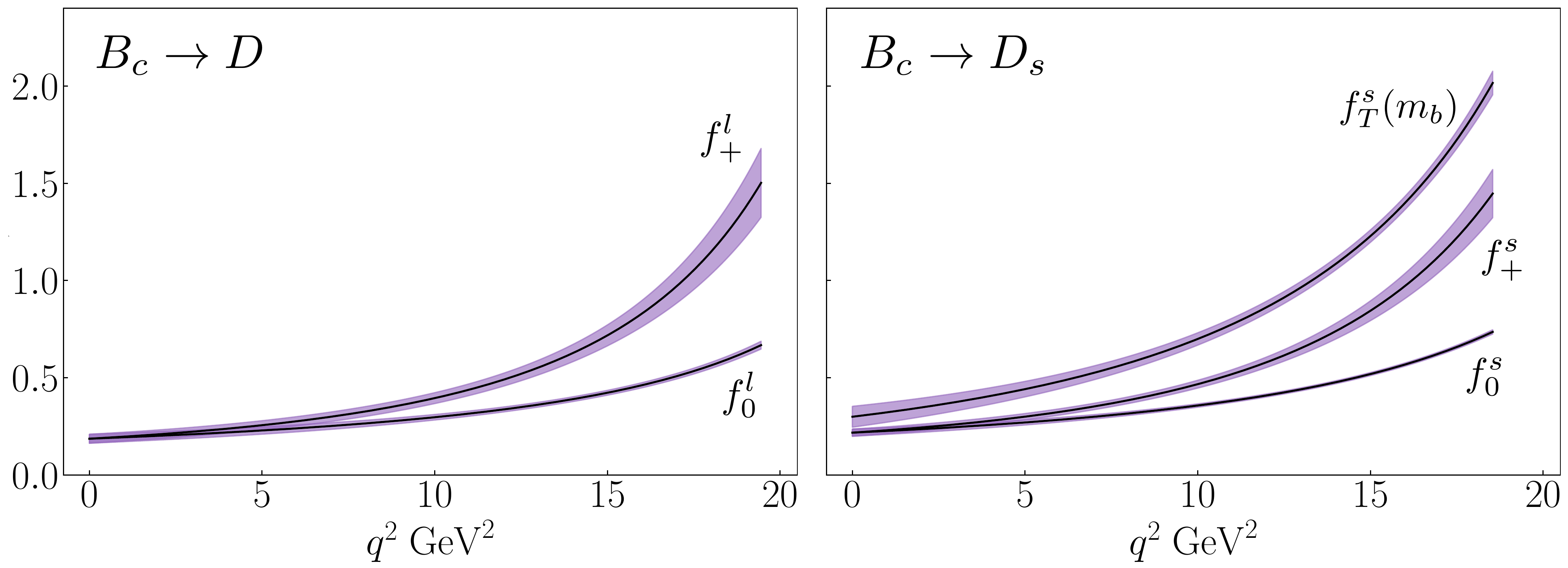}
	\caption{The fit functions for the $B_c \to D_l$ (left) and $B_c \to D_s$ (right) form factors $f_{0,+}$ and $f_{0,+,T}$ respectively tuned to the continuum limit with physical quark masses.}
	\label{fig:final_ffs}
\end{figure}

In Fig.~\ref{fig:final_ffs}, we present our form factors in the limit of vanishing lattice spacing and physical quark masses across the entire physical range of $q^2$.
Details of the results of the fits of correlation functions and lattice form factors from which Fig.~\ref{fig:final_ffs} is derived from are given in~\cite{Cooper:2021ofu}.
%

\subsection{Observables for $B_c^+ \to D^0 \ell^+ \nu_{\ell}$} \label{sec:BcD_obs}

\begin{figure}
	\centering
	\includegraphics[width=0.55\textwidth]{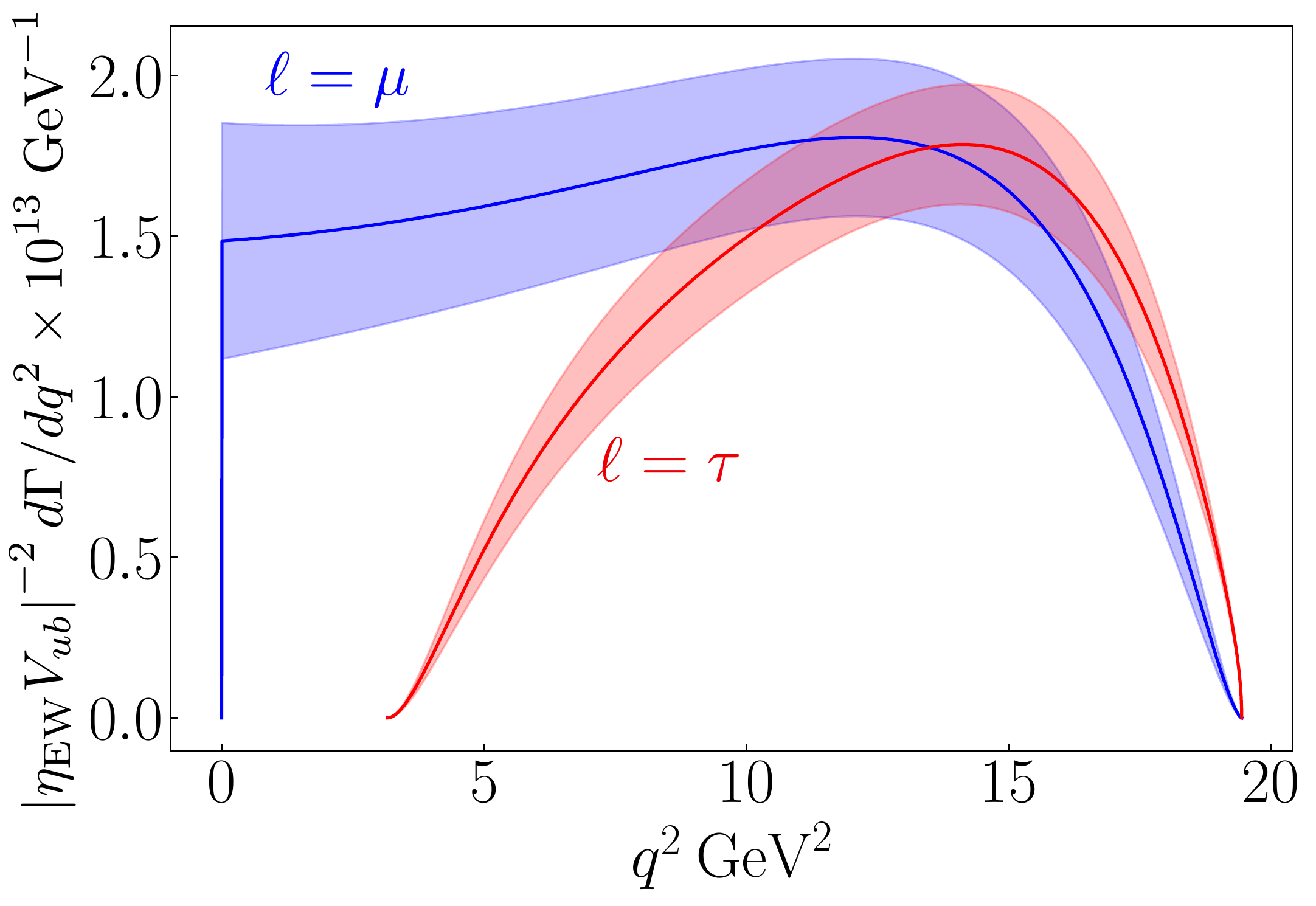}
	\caption{The differential decay rate $\eta_{\mathrm{EW}}^{-2}|V^{ub}|^{-2} d \Gamma (B_c^+ \to D^0 \ell^+ \nu_{\ell}) / dq^2$ as a function of $q^2$ for the cases $\ell = \mu$ in blue and  $\ell = \tau$ in red.}
	\label{fig:diff_decay_rate_BcD}
\end{figure}

We plot the differential decay rate $\eta_{\mathrm{EW}}^{-2} |V_{ub}|^{-2} d \Gamma (B_c^+ \to D^0 \ell^+ \nu_{\ell}) / dq^2$ derived from our form factors as a function of $q^2$ in Fig.~\ref{fig:diff_decay_rate_BcD}.
%
%
We integrate this function to find $\eta_{\mathrm{EW}}^{-2} |V^{ub}|^{-2} \Gamma$.
This is then combined with $\eta_{\mathrm{EW}}$, the CKM matrix element $V_{ub} = 3.82(24) \times 10^{-3}$~\cite{ParticleDataGroup:2020ssz} (an average of inclusive and exclusive determinations), and the lifetime of the $B_c$ meson~\cite{Aaij:2014gka} to obtain the branching ratios in Table~\ref{tab:integrated_Gamma_BcDl}.
%
\begin{table}
	\centering
	\caption{For $B_c^+ \to D^0 \ell^+ \nu_{\ell}$, we give values for the branching ratios (BR) for each of the cases $\ell = e,\mu,\tau$.
		We take the lifetime of $B_c$ meson to be $513.49(12.4) \: \mathrm{fs}$~\cite{Aaij:2014gka}.
		The errors from the lifetime and the CKM matrix element $V_{ub}$ are shown explicitly.
		The error from $\eta_{\mathrm{EW}}$ is negligible.
		We ignore uncertainties from long-distance QED contributions since the meson $D^0$ in the final state is neutral.}
	\begin{tabular}{c | c}
		\hline\hline
		decay mode & BR$\times 10^{5}$ \\ [0.1ex]
		\hline
		$B_c^+ \to D^0 e^+ \nu_e$  & $3.37(48)_{\mathrm{lattice}}(8)_{\tau_{B_c}} (42)_{\mathrm{CKM}}$ \\
		$B_c^+ \to D^0 \mu^+ \nu_\mu$  & $3.36(47)_{\mathrm{lattice}}(8)_{\tau_{B_c}}(42)_{\mathrm{CKM}} $  \\
		$B_c^+ \to D^0 \tau^+ \nu_\tau$  & $2.29(23)_{\mathrm{lattice}}(6)_{\tau_{B_c}}(29)_{\mathrm{CKM}} $  \\
		\hline\hline
	\end{tabular}
	\label{tab:integrated_Gamma_BcDl}
\end{table}
%
%
At present, errors from our lattice calculation dominate those associated with the lifetime of the $B_c$ meson and the CKM matrix element $V_{ub}$.

We compare our results with those for the decay mode $B_{c}^{+} \rightarrow J / \psi \mu^{+} \nu_{\mu}$.
We take the form factors for this decay from HPQCD's lattice QCD calculation in~\cite{Harrison:2020gvo}.
We combine these form factors with those for $B_{c}^{+} \rightarrow D^{0} \ell^{+} \nu_{\ell}$ computed in this study to find the ratios
%
\begin{align}
	\left|\frac{V_{cb}}{V_{ub}} \right|^2 \frac{\Gamma( B_c^+ \to D^0 \mu^+ \nu_\mu )}{\Gamma(B_{c}^{+} \rightarrow J / \psi \mu^{+} \nu_{\mu})} = 0.257(36)(18), \nonumber \\
	\left|\frac{V_{cb}}{V_{ub}} \right|^2 \frac{\Gamma( B_c^+ \to D^0 \tau^+ \nu_\tau )}{\Gamma(B_{c}^{+} \rightarrow J / \psi \tau^{+} \nu_{\tau})} = 0.678(69)(45).
\end{align}
The first error comes from our form factors for $B_{c}^{+} \rightarrow D^0 \mu^+ \nu_{\mu}$, and the second error comes from the form factors for $B_{c}^{+} \rightarrow J / \psi \mu^{+} \nu_{\mu}$ in~\cite{Harrison:2020gvo}.
We (conservatively) treat the form factors for $B_{c}^{+} \rightarrow J / \psi \mu^{+} \nu_{\mu}$ as uncorrelated to the $B_{c}^{+} \rightarrow D^{0} \ell^{+} \nu_{\ell}$ form factors.

\subsection{Observables for $B_c^+ \to D_s^+ \ell^+ \ell^-$} \label{sec:BcDs_obs}

Like $B \to K \ell^+ \ell^-$, the process $B_c^+ \to D_s^+ \ell^+ \ell^-$ is a rare decay mediated by the loop-induced $b \to s$ transition.
Here, we follow nomenclature commonly used for $B \to K \ell^+ \ell^-$ as in~\cite{Becirevic:2012fy} and replace the initial and final mesons in the $B \to K$ formulae with $B_c$ and $D_s$ respectively.
We calculate observables for $B_c^+ \to D_s^+ \ell^+ \ell^-$ from our form factors $f_{0,+,T}^s$ ignoring small non-factorisable contributions at low $q^2$~\cite{Khodjamirian:2012rm, Hambrock:2015wka}.

\begin{figure}
	\centering
	\includegraphics[width=1.0\textwidth]{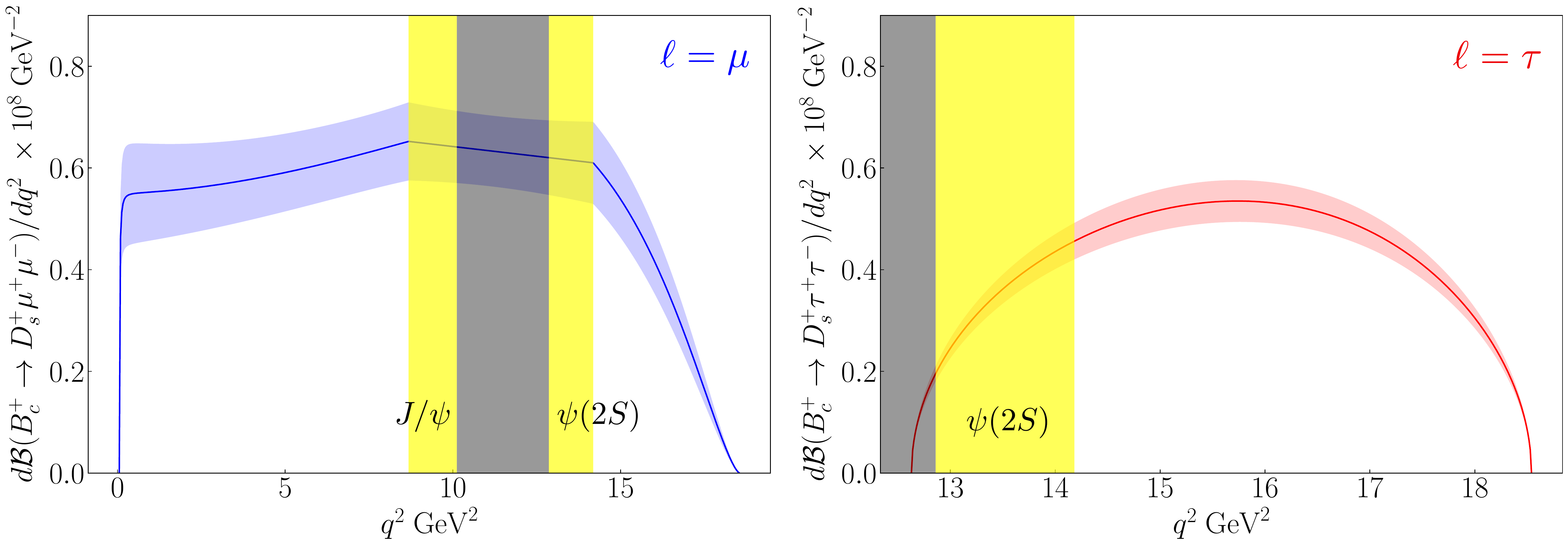}
	\caption{Plot of the $B_c^+ \to D_s^+ \ell^+ \ell^-$ differential branching ratio for $\ell = \mu$ (top) and $\ell = \tau$ (bottom) in the final state. The yellow bands show regions where charmonium resonances (not included in our calculation) could have an impact. The grey band is between the two yellow regions labelling the charmonium resonances. Through the yellow and grey bands, we interpolate the function $d\mathcal{B}_{\mu}/dq^2$ linearly when integrating to find the branching fraction and related quantities.} 
	\label{fig:dbdqsqbcdslepton}
\end{figure}

In Fig.~\ref{fig:dbdqsqbcdslepton} we plot the differential branching fractions for the cases $\ell = \mu, \tau$ for the physical range $4m_{\ell}^2 < q^2 < (M_{B_c} - M_{D_s})^2$.
These are constructed from the expressions in~\cite{Becirevic:2012fy} for $B\to K$.
The yellow bands span across $\sqrt{q^2} = 2.956-3.181 \; \mathrm{GeV}$ and $3.586-3.766 \; \mathrm{GeV}$.
These regions are the same as in~\cite{Aaij:2012cq} and they represent veto regions which largely remove contributions from charmonium resonances via intermediate $J/\psi$ and $\psi(2S)$ states.
The effects of charmonium resonances are not included in our differential branching fractions.
For $d \mathcal{B}_{\mu} / dq^2$  between $\sqrt{q^2} = 2.956$ and $\sqrt{q^2} = 3.766$, we interpolate the function linearly as done in~\cite{Du:2015tda} for the $B \to K$ branching fraction.

On integrating with respect to $q^2$, we report on the ratio
\begin{align}
	R_{\ell_2}^{\ell_1}\left(q_{\text {low }}^{2}, q_{\text {high }}^{2}\right) = \frac{\int_{q_{\text {low }}^{2}}^{q_{\text {high }}^{2}} d q^{2} d \mathcal{B}_{\ell_1} / d q^{2}}{\int_{q_{\text {low }}^{2}}^{q^2_{\text {high }}}{d q^{2}} d \mathcal{B}_{\ell_2} / d q^{2}}
\end{align}
for different choices of final-state lepton $\ell_{1,2}$ and integration limits $q^2_{\mathrm{low}}, q^2_{\mathrm{high}}$.
We find that
\begin{align}
	R_{e}^{\mu}\left(4 m_{\mu}^{2}, q_{\max }^{2}\right) &=1.00203(47) \\
	R_{e}^{\mu}\left(1 \; \mathrm{GeV}^2, 6 \; \mathrm{GeV}^2\right) &= 1.00157(52) \label{eqn:mu_e_lower_ratio}\\
	R_{e}^{\mu}\left(14.18 \;\mathrm{GeV}^{2}, q_{\max }^{2}\right) &= 1.0064(12) \\
	R_{e}^{\tau}\left(14.18 \;\mathrm{GeV}^{2}, q_{\max }^{2}\right) &= 1.34(13) \\
	R_{\mu}^{\tau}\left(14.18 \; \mathrm{GeV}^{2}, q_{\max }^{2}\right) &= 1.33(13)
\end{align}
The latter two ratios above involve the differential decay widths above the veto region associated with the resonance from $\psi(2S)$.
Over the range $4m_{\ell}^2 < q^2 < q^2_{\mathrm{max}}$, we obtain the branching fractions given in Table~\ref{tab:integrated_Gamma_BcDs}.
%
\begin{table}
	\centering
	\caption{For $B_c^+ \to D_s^+ \ell^+ \ell^-$, we give values for $d\mathcal{B}/dq^2 \times 10^{7}$ integrated with respect to $q^2$ over the given ranges $(q^2_{\mathrm{low}}, q^2_{\mathrm{high}})$ in $\mathrm{GeV}^2$ for each of the cases $\ell = e,\mu,\tau$.
		We take the lifetime of $B_c$ meson to be $513.49(12.4) \: \mathrm{fs}$~\cite{Aaij:2014gka}. Note that these results do not include effects from charmonium or $u \overline{u}$ resonances. }
	\begin{tabular}{c | c c c }
		\hline\hline
		decay mode & $(4 m_{\ell}^2, q^2_{\mathrm{max}})$ & $(1, 6)$ & $(14.18, q^2_{\mathrm{max}})$  \\ [0.1ex]
		\hline
		$B_c^+ \to D_s^+ e^+ e^-$ & $1.00(11)$ & $0.285(41)$ & $0.146(22)$\\
		$B_c^+ \to D_s^+ \mu^+ \mu^-$ & $1.00(11)$ & $0.286(41)$ & $0.147(22)$ \\
		$B_c^+ \to D_s^+ \tau^+ \tau^-$ & $0.245(18)$ & --- & $0.195(14)$ \\
		\hline\hline
	\end{tabular}
	\label{tab:integrated_Gamma_BcDs}
\end{table}
%

%

\section{Conclusions and Outlook}

For the first time from lattice QCD, we obtain the scalar and vector form factors $f_{0,+}$ for $B_c \to D_l$ and the scalar, vector and tensor form factors $f_{0,+,T}$ for $B_c \to D_s$ across the entire physical ranges of $q^2$ in the continuum limit with physical quark masses.
Our lattice QCD calculation uses four different lattices with three different lattice spacings, unphysical and physically massive light quarks, and a range of heavy quark masses.
Together the lattice data informs the limit of vanishing lattice spacing, physical $b$ quark mass, and physical (equal-mass) up and down quark masses.

The error on the decay widths $\Gamma(B_c^+ \to D^0 \ell^+ \nu_{\ell})$ (see Table~\ref{tab:integrated_Gamma_BcDl}) from our form factors is similar to the error on the present determination of $V_{ub}$.
For the cases $\ell = e$ or $\mu$, the lattice error is $13\%$ larger than the error from $V_{ub}$, whereas, for $\ell = \tau$, the lattice error is nearly $20\%$ smaller than the error from $V_{ub}$.
The error on the form factors calculated here for $B_c \to D_s$ is smaller than that for $B_c \to D$ by up to a factor of 2 at small recoil.

Experimental observations are expected from LHC in the near future~\cite{Bediaga:2018lhg}.
In Sections~\ref{sec:BcD_obs} and~\ref{sec:BcDs_obs} we give results for a host of observables that can be compared to experiment.
In~\cite{Cooper:2021ofu}, we demonstrate how the uncertainties in our calculation can be reduced in the future to complement experimental results as they improve. We consider a finer lattice on which we simulate directly at the mass of the $b$ quark, and an alternative extraction of $f_+$ from lattice matrix elements of the spatial vector current to improve the uncertainty of $f_+$ near zero-recoil.


\section*{Acknowledgments}

We thank Jonna Koponen, Andrew Lytle, William Parrott and Andre Zimermmane-Santos for making previously generated lattice propagators available for our use; we thank Daniel Hatton et al. for the calculation of $Z_T$ in~\cite{Hatton:2020vzp}, and Chris Bouchard, Judd Harrison and William Parrott for useful discussions.
We are grateful to the MILC collaboration for making publicly available their gauge configurations and their code MILC-7.7.11 \cite{MILCgithub}.
This work was performed using the Cambridge Service for Data Driven Discovery (CSD3), part of which is operated by the University of Cambridge Research Computing on behalf of the STFC DiRAC HPC Facility.
The DiRAC component of CSD3 was funded by BEIS capital funding via STFC capital grants ST/P002307/1 and ST/R002452/1 and STFC operations grant ST/R00689X/1.
DiRAC is part of the National e-Infrastructure.
We are grateful to the CSD3 support staff for assistance.
This work has been partially supported by STFC consolidated grant ST/P000681/1.

\bibliographystyle{JHEP}
\bibliography{bib}
%

\end{document}